# CFD-Based Quantification of Hemodynamic Variables in Cerebral Aneurysms: How Hemodynamics Shape Aneurysm Fate

Reza Bozorgpour*[1], Pilwan Kim[1]

[1]Department of Biomedical Engineering, University of Wisconsin-Milwaukee, Milwaukee, WI, USA

*Corresponding author

E-mail: bozorgp2@uwm.edu

## Abstract

Cerebral aneurysms are pathological dilations of intracranial arteries that can rupture with devastating consequences, including subarachnoid hemorrhage, stroke, and death. Accumulating evidence indicates that local hemodynamic forces play a critical role in aneurysm initiation, growth, and rupture. Computational fluid dynamics (CFD) and imaging-based techniques have enabled the extraction of various hemodynamic variables to characterize these flow conditions. However, the literature is highly fragmented, with different studies adopting distinct sets of metrics—such as wall shear stress (WSS), oscillatory shear index (OSI), wall shear stress gradient (WSSG), relative residence time (RRT), or endothelial cell activation potential (ECAP)—making it difficult to compare results or establish standardized methodologies. This paper provides the first comprehensive catalog of hemodynamic variables used in cerebral aneurysm studies to date. By systematically identifying and organizing these parameters based on their physical basis and frequency of use, this work offers a consolidated reference to guide future research. The goal is to support consistent variable selection, enhance reproducibility, and facilitate the design of more robust studies linking vascular biomechanics to aneurysm pathophysiology. This review aims to serve as a foundational resource for researchers and clinicians seeking to incorporate hemodynamic modeling into cerebral aneurysm analysis and risk assessment.

# 1. Introduction

Cerebral aneurysms result from structural vulnerabilities in the walls of arteries supplying the brain, predisposing these regions to pathological dilation and potential rupture. These lesions vary significantly in dimension, from under 0.5 mm to over 25 mm in diameter. The saccular subtype, which constitutes the majority, often shows degeneration or loss of the tunica media and fragmentation of the internal elastic lamina [1]. Less frequent forms, such as fusiform and mycotic aneurysms, also occur but are less well characterized. Frequently asymptomatic, these aneurysms are commonly discovered through advanced neuroimaging techniques or post-mortem analysis. Around 85% are situated within the anterior circulation, typically at points where vessels bifurcate. Key contributors to their development and enlargement include disturbed hemodynamics, chronic hypertension, traumatic injury, atherosclerotic changes, and inherited predispositions [2].

Subarachnoid hemorrhage (SAH), a severe outcome of aneurysm rupture, affects approximately 10 out of every 100,000 individuals annually [3] and is associated with high rates of disability and death [4]. Although modern imaging has improved the ability to detect unruptured aneurysms, tools for reliably assessing rupture risk have not advanced at the same pace [4-6].

The risk of cerebral aneurysm rupture arises from a multifactorial interaction between inherent and external influences. Among the nonmodifiable contributors are age, biological sex, and inherited genetic factors. Epidemiological data indicate that aneurysms are more commonly observed in older adults, women, and individuals with a family history of the condition. In contrast, modifiable risk factors pertain to behavioral and medical conditions, such as chronic hypertension, tobacco use, and high alcohol intake. Elevated blood pressure increases the mechanical load on arterial walls, intensifying vascular stress. Meanwhile, habits like smoking and excessive drinking promote vascular inflammation and degeneration, thereby facilitating aneurysm development and progression [7-12].

Identifying individuals at high risk for cerebral aneurysm rupture is essential to prevent severe clinical consequences. Although aneurysms can develop in anyone, certain demographic and medical factors—such as older age, female sex, family history, hypertension, and arteriosclerosis—are linked to higher susceptibility [13-16]. Beyond patient characteristics, the aneurysm's shape, location, and local blood flow patterns also influence rupture potential [17-20]. Structurally, the aneurysm dome often lacks key elastic and muscular components, weakening the vessel wall and making it vulnerable to rupture under hemodynamic stress [7, 11, 21].

Recent research has increasingly emphasized the importance of hemodynamic forces in the development and rupture of cerebral aneurysms [19, 22]. Hemodynamics, the study of blood flow and its interaction with vessel walls, is now seen as a central factor in aneurysm pathophysiology. Among the most studied variables is wall shear stress (WSS), a tangential force exerted by blood flow. Elevated WSS has been linked to endothelial injury and inflammation, promoting aneurysm initiation, while reduced WSS is associated with structural weakening and rupture risk [22, 23]. Other parameters, including pressure, velocity, and flow patterns, have also been explored for their contributions to aneurysm behavior [24, 25]. Computational fluid dynamics (CFD) has been widely used to model these interactions and study how blood flow dynamics relate to aneurysm morphology [26, 27].

Treatment decisions, whether surgical or endovascular, carry inherent risks. Clinicians aim to discern which aneurysms are stable, which pose a future threat, and which warrant proactive intervention [5]. This has led to the development of rupture prediction models that incorporate risk factors such as lesion size and anatomical location, especially within the posterior cerebral circulation [6].

Over the past two decades, increased understanding of intra-aneurysmal blood flow has sharpened our ability to assess rupture potential. Computational simulations are now widely used in clinical research to analyze hemodynamic forces acting on aneurysm walls. This review discusses key geometric features and flow-related parameters linked to aneurysm formation, progression, and rupture risk.

## 2. Morphological Determinants of Aneurysm Rupture

**Size:** Geometric characteristics of aneurysms, particularly their size and shape, are central to clinical decision-making [28, 29]. Unlike dynamic hemodynamic parameters, these morphological features remain constant and are easier to quantify. Even small alterations in aneurysm geometry can significantly influence internal blood flow patterns, highlighting their relevance in assessing rupture risk [30]. Numerous studies have identified aneurysm size as a predictor of both formation and rupture likelihood [31-33]. In a study of 118 patients, Korja et al. [15] found that aneurysms exceeding 7 mm in diameter were associated with a notably higher lifetime risk of SAH, despite a median aneurysm size of just 4 mm (ranging from 2 to 25 mm). The assumption that bigger aneurysms are more prone to rupture is widespread, but the exact point at which size becomes dangerous is still under debate. While some researchers flag diameters beyond 7 or 10 mm as particularly hazardous, others caution that even small aneurysms shouldn't be dismissed as harmless [34, 35]. Just as important as size is how fast an aneurysm grows, rapid enlargement can be a red flag, often signaling a higher chance of rupture.

**Shape:** Aneurysm morphology plays a role in rupture assessment, but shape alone does not offer a reliable basis for prediction [36]. In attempts to better characterize saccular aneurysms, some researchers have proposed subclassifications—such as elliptical, lobulated, beehive, and pear-shaped forms. Among these, elliptical aneurysms appear more frequently in ruptured cases, while rounded shapes are commonly seen in unruptured ones [37]. Yet despite such distinctions, the link between aneurysm geometry and rupture remains a subject of debate, with evidence suggesting that anatomical features must be considered alongside patient-specific clinical data. Moreover, aneurysms exhibiting non-uniform or complex geometries, such as fusiform or multilobular configurations, have been associated with higher rupture potential, likely due to structural fragility or uneven distribution of wall stress [8, 24, 38, 39]. These shape irregularities may indicate biomechanical instability, signaling regions where the vessel wall is more susceptible to rupture under fluctuating hemodynamic loads.

To improve the accuracy of rupture risk evaluation, researchers have introduced several geometric parameters that extend beyond basic aneurysm size measurements [15, 32, 40]. Two of the most used shape-based indicators are the aspect ratio (AR) and the size ratio (SR), as depicted in Figure 1 [41].

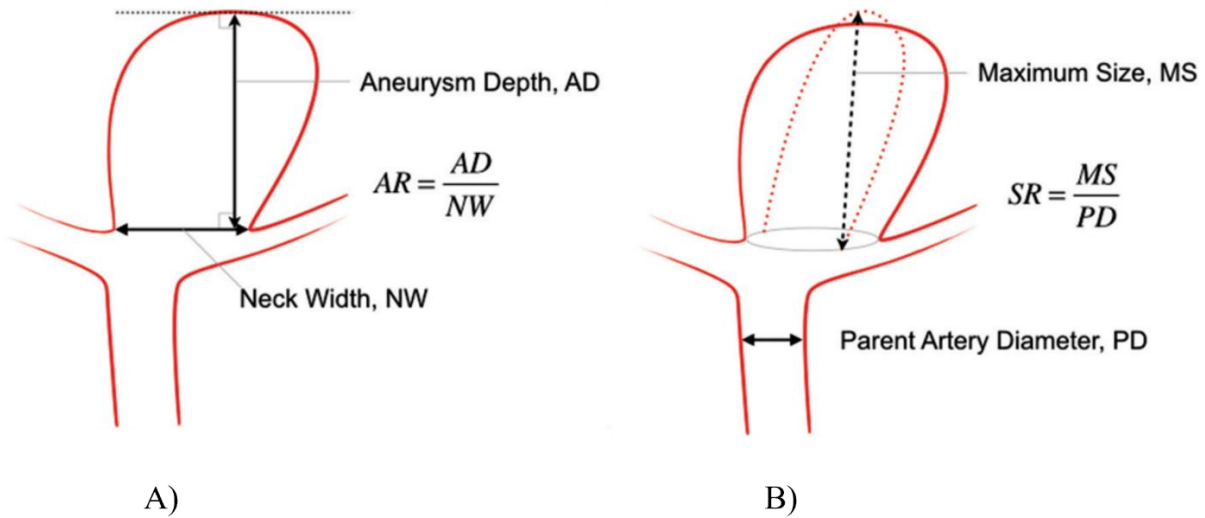

Figure 1: (A) Aspect ratio. (B) Size ratio

Aneurysms with tall domes and narrow necks, quantified by a high AR, are believed to present a greater risk of rupture. AR is calculated by dividing the vertical height of the aneurysm by the width of its neck. When this ratio increases, the aneurysm tends to trap blood within the sac, disrupting normal flow and creating zones of slow-moving blood. These low-flow environments are often marked by reduced WSS, which is a mechanical stimulus essential for maintaining endothelial health. In such conditions, inflammatory activity tends to rise within the vessel wall, potentially weakening its structure. This vulnerability is especially pronounced in aneurysms with deeper domes and AR values at or above 1.5, a threshold frequently cited as a strong predictor of rupture risk [42-46]. Ujiie et al. [47] showed that ruptured aneurysms had a higher mean AR of 2.7 compared to 1.8 in unruptured ones. Similarly, Jiang et al. [48] found that 80% of ruptured aneurysms had ARs above 1.6, while 90% of unruptured cases were below this threshold. In another study of 119 ruptured aneurysms across key brain arteries, median AR values ranged from 0.92 to 1.13 [41].

The SR, calculated by comparing aneurysm height to the diameter of the parent artery, serves as a valuable morphological indicator [8, 20, 24, 38]. A higher SR suggests disproportionate aneurysm growth and has been linked to elevated rupture risk due to increased wall tension. Studies have shown that ruptured aneurysms often display significantly higher SR values, with a threshold of 2.3 proposed as a predictive marker, even for small aneurysms under 5 mm in diameter [44]. While promising, this cutoff may vary depending on the studied population and analytical approach.

**Location**: is also a key factor in aneurysm rupture risk. Aneurysms situated at arterial bifurcations or terminations face higher rupture potential due to concentrated hemodynamic stress [8, 49, 50]. Regions like the posterior communicating artery and middle cerebral artery are particularly prone to rupture and are considered high-risk sites.

Table 1 lists commonly used morphological parameters in cerebral aneurysm studies, along with their definitions, to quantify aneurysm geometry for clinical and computational analysis.

Table 1: Different morphological parameters of clinical interest in aneurysms.

| Morphological parameter | Definitions |
| --- | --- |
| Maximum Size/Dome Width (DW) [51-54] | Maximum aneurysm diameter measured along the plane parallel to the neck |
| Aspect Ratio (AR) [52, 55] | Ratio of aneurysm dome height to the width of its neck |
| Size Ratio (SR) [56-58] | Ratio of the aneurysm's maximum dimension to the diameter of the parent vessel |
| Maximum Dome Height (MDH) [55, 59] | Distance from the center of the neck plane to the apex of the aneurysm dome |
| Neck Width (NW) [52, 53] | Mean diameter of the aneurysm's neck plane |
| Dome Perpendicular Height (DPH) [55, 59] | Perpendicular distance from neck center to aneurysm apex |
| Parent Vessel Diameter (PVD) [53, 55] | Inlet artery diameter of the aneurysm |
| Bottle Neck Ratio (BNR) [55, 59] | Ratio of aneurysm's maximum dimension to neck width |
| Inflow angle (IFA) [52, 56] | Angle between the inlet centerline and aneurysm axis |
| Undulation Index (UI) [45, 53, 58] | Measure of the aneurysm's surface undulation intensity |
| Non-Sphericity Index (NSI) [45, 53, 55, 56, 58] | Quantifies how much the aneurysm shape deviates from a perfect sphere |
| Ellipticity Index (EI) [45, 53, 58] | Metric comparing the ellipticity of the aneurysm geometry |
| Natural Frequency ($\omega$) [60] | A metric that uses an object's natural frequency to assess variations in its shape and size |

## 3. Flow-Related Factors in Cerebral Aneurysm Behavior

Understanding the forces that blood flow exerts on the vessel wall is central to grasping how cerebral aneurysms form, expand, and eventually rupture. These mechanical forces, collectively studied under the field of hemodynamics, are not only essential for maintaining vascular homeostasis but also play a pivotal role in pathological remodeling when disrupted. The dynamic interaction between blood and vascular structures influences key biological processes such as endothelial integrity, inflammatory signaling, and vessel wall tension [25, 61-63]. Hemodynamic forces include multiple components, each with distinct mechanical implications. Shear stress, which acts tangentially to the inner wall of the vessel, arises from frictional contact between blood and endothelium. Normal stress, on the other hand, results from blood pressure pushing perpendicularly against the vessel wall, while circumferential tensile stress stretches the wall outward along its radius. These stress patterns, often visualized in simplified straight-vessel models, collectively influence vascular remodeling and may predispose certain regions to aneurysmal transformation. In recent years, growing attention has been directed toward identifying which of these flow-induced forces most strongly contribute to aneurysm destabilization. The integration of hemodynamic insights into clinical and computational models has therefore become essential for predicting aneurysm behavior and rupture risk.

### 3.1. Wall Shear Stress (WSS)

Within the vascular system, communication between blood flow and vessel walls is constant and dynamic. One of the most influential forces in this dialogue is WSS, a subtle, tangential force generated as blood slides along the inner lining of arteries [22, 23, 64, 65]. Under balanced conditions, WSS serves as a stabilizing cue, guiding endothelial behavior, maintaining anti-inflammatory signaling, and promoting vascular homeostasis. But when this mechanical signal shifts, either becoming excessively forceful or dissipating into near stagnation, the consequences are biologically disruptive. Elevated WSS can act as a blunt instrument, stripping away endothelial layers and provoking inflammatory responses that support aneurysm growth. In contrast, low WSS fosters chaotic, low-velocity flow patterns where endothelial cells falter, immune cells accumulate, and the surrounding matrix begins to degrade. These environments, whether under excessive stress or deprived of flow, become fertile ground for the birth and eventual rupture of cerebral aneurysms. For many years, researchers have examined the role of WSS in the initiation, growth, and rupture of cerebral aneurysm; WSS is a dynamic force induced by the movement of a viscous fluid along the surface of blood vessel wall [66]. Generally, WSS is considered a primary parameter in cerebral aneurysm hemodynamics.

Aneurysm wall tissue is subjected to dynamic and evolving flow conditions over time. Early bleb formation is often linked to high WSS, which may transition to low WSS as aneurysm geometry changes [64]. Both high and low WSS have been implicated in aneurysm progression and rupture. Notably, blebs tend to form near regions of elevated WSS rather than areas of stagnation [67]. While Jou et al. [68] found no significant difference in mean WSS between ruptured and unruptured aneurysms, ruptured cases showed a larger surface area exposed to low WSS. Supporting this, Xiang et al. [58] also reported an association between low WSS and rupture, and Boussel et al. [42] observed that aneurysm growth often occurs in low-WSS zones. In unruptured aneurysms, thin-walled dome regions have been spatially linked to areas of reduced wall stress [69]. Cebral et al. [70] identified similarities between flow in hyperplastic aneurysm regions and that in atherosclerotic plaques, characterized by low and recirculating WSS. Moreover, Doddasomayajula et al. [71] using mirrored aneurysms from the same patients, showed that ruptured aneurysms exhibited more extreme WSS oscillations than unruptured ones.

### 3.1.1. Shear Stress and Endothelial Cell Responses

Endothelial cells (ECs), which line the vessel lumen, rely on WSS for normal function and are among the first to respond to hemodynamic changes. Abnormal WSS can disrupt EC integrity, trigger inflammation via mechanosensors (e.g., ion channels, integrins, GPCRs), and activate pathways like NF-κB [72]. Loss of WSS, such as that caused by thrombus formation, further dysregulates ECs [25,38]. Under laminar flow, ECs maintain homeostasis and align with the direction of flow, limiting thrombosis and inflammation [42, 72]. In contrast, disturbed or low WSS accelerates EC turnover, promotes apoptosis [34,37], and shifts EC secretion profiles toward inflammatory and vasoconstrictive factors. Low WSS also facilitates leukocyte adhesion via selectins [65], and lymphocytes have been implicated in aneurysm formation and rupture [73]. Studies show that chronic high WSS increases MMP expression, degrading the extracellular matrix (ECM) and promoting vascular remodeling and growth factor production [74]. As aneurysm diameter increases, WSS decreases, lowering matrix degradation signals. ECs also regulate smooth muscle cells (SMCs) through NO and prostaglandins; low WSS reduces NO and PGI2, promoting oxidative stress, atherosclerosis, and wall degeneration [75-81]. Inflammatory activation of NF-κB by IL-1β and TNF-α enhances proinflammatory mediators (e.g., COX-2, PGE2) and proteases [75, 82, 83]. Aoki et al. [84, 85] showed that inhibiting COX-2 or EP2 reduces aneurysm formation and inflammation, suggesting a key role for this pathway. Aspirin, a COX-2 inhibitor, has been associated with reduced aneurysm rupture risk [86]. WSSG also influences EC behavior. Additionally, damage to the internal elastic lamina (IEL), which separates ECs from the media, is linked to increased MMP-2/9 secretion by SMCs after endothelial injury [78]. Table 2 outlines how different hemodynamic patterns influence vascular wall biology based on recent findings.

**Table 2: Summary of hemodynamics pattern and vascular wall changes**

| Authors & Year | Hemodynamics Pattern | Vascular Wall Changes |
|---|---|---|
| Li et al. [87], 2025 | Low WSS | Triggers endothelial apoptosis and dysfunction via the miR-330/SOD2/HSP70 pathway, promoting atherosclerosis progression. |
| LV et al. [88], 2024 | | Promotes endothelial pyroptosis and atherosclerosis via the IKKε/STAT1/NLRP3 inflammatory pathway |
| Chen et al. [89], 2024 | | Increased endothelial permeability via junctional disruption under low or oscillatory shear stress |
| Kumar et al. [90], 2018 | | Low WSS associated with increased vasoconstriction and severe endothelial dysfunction in coronary artery disease |
| Zhou et al. [65], 2017 | | Enhanced leukocyte rolling driven by selectin activation |
| Lu & Kassab [79], 2011 | | Reduced $PGI_2$ production promoting atherosclerosis and heightened oxidative stress |
| Boussel et al. [42], 2008 | | Endothelial disorganization, increased apoptosis, and upregulation of vasoconstrictive and inflammatory mediators |
| Papadaki et al. [91], 1998 | | Reduced production of tPA |
| Simões-Faria et al. [92], 2025 | High WSS | Shifts endothelial metabolism by enhancing glutamine utilization and suppressing glycolysis, promoting metabolic adaptation for vascular homeostasis. |
| Lu et al. [93], 2024 | | Induces endothelial podosome formation and collagen IV degradation, promoting intracranial aneurysm development in a novel mouse model. |
| Hiroshima et al [94], 2022 | | Alters MMP/TIMP expression in an EC–SMC coculture model, highlighting mechanosensitive vascular remodeling. |
| Pawlowska et al.[82], 2018 | | Elevates IL-1β expression, contributing to vascular inflammation and aneurysm progression |
| Taylor et al.[95], 2015 | | Drives SMC migration to the intima, induces phenotypic shift to secretory type, and increases MMP production |
| Fukuda & Aoki [96], 2015 | | Upregulates NF-κB and COX-2, enhancing inflammatory signaling in endothelial cells |
| Shi & Tarbell [97], 2011 | | Induces SMCs to release PDGF and FGF-2, promoting cell migration and vascular remodeling |
| Dhawan et al. [98], 2007 | | Alters endothelial phenotype and promotes ECM disruption, contributing to plaque development. |
| Papadaki et al.[99], 1998 | | Increased production of tPA |
| Dolan et al. [100], 2013 | Negative WSSG | Suppresses gene expression typically induced by high WSS. |
| | Positive WSSG | Downregulates anti-apoptotic and inflammatory genes, while upregulating ECM-degrading proteases like ADAMTS1 and reducing TAGLN, a repressor of MMP-9 |
| Turjman et al. [72], 2014 | Laminar Flow | Promotes endothelial cell alignment with flow direction, reducing thrombosis and leukocyte adhesion. |
| | Turbulent Flow | Disrupts endothelial organization, induces cuboidal cell shape, and enhances thrombosis and leukocyte adhesion. |

### 3.1.2. Shear Stress and SMC Phenotype Changes

SMCs, typically shielded from WSS, respond to increased pressure and wall tension through contraction, proliferation, matrix synthesis, and phenotypic switching [97]. Under stress, they transition from a contractile to a proinflammatory phenotype, producing NF-κB, IL-1β, TNF-α, and MMPs, and undergo apoptosis [95, 101, 102]. IL-1β promotes SMC apoptosis during aneurysm progression; although knockout mice formed aneurysms at similar rates, they showed fewer advanced lesions and reduced apoptosis [75]. While macrophages produce MMP-2/9 [103], SMCs were also shown to express these enzymes after flow-induced injury, even in the absence of inflammatory cells [104]. These SMCs showed reduced contractile markers (SMA, calponin) and increased NF-κB and MCP-1 expression. SMCs release PDGF and FGF-2 in response to shear stress, promoting migration, reducing SMA, and disrupting ECM structure [37,39]. Receptors for TGF-β, PDGF, and VEGF are upregulated in ruptured aneurysms and linked to wall remodeling and "warning leaks" [81, 105]. Ets-1 expression, which regulates MCP-1, is also induced during aneurysm development [106]. High WSS increases tPA and PAR-1 expression, while low WSS reduces them [91]. Ruptured aneurysms are more often associated with luminal thrombosis [107]. Frösen et al. [97] also showed that disorganized SMC alignment—modulated by shear stress uniformity—correlates with higher rupture risk.

Under pulsatile blood flow conditions, the magnitude of WSS at each point along the vessel wall varies throughout the cardiac cycle. To capture its overall effect, the TAWSS is often used as a representative metric across one full cycle [58, 108]. This value is computed using the following equation:

$$TAWSS = \frac{1}{T}\int_0^T |\tau_w| \qquad \text{(Eq. 1)}$$

In this formulation, $\tau_w$ represents the instantaneous shear stress vector at a given point on the vessel wall, while $T$ denotes the duration of one complete cardiac cycle [109, 110]. TAWSS reflects the average magnitude of shear forces acting throughout the cycle, computed at each mesh point on the vessel surface. Expressed in Pascals, typical TAWSS values range from approximately 1.5 to 10 Pa, offering insight into sustained mechanical loading conditions across the cardiac phase.

### 3.2. WSS gradient (WSSG)

To evaluate the spatial variation of shear forces along the vessel wall, researchers computed the Wall Shear Stress Gradient (WSSG) [111] using the spatial derivative of WSS. Beyond this primary metric, two additional indicators were derived: the absolute WSSG (absWSSG), which captures magnitude regardless of direction, and the directional WSSG (dirWSSG), which distinguishes vector alignment. For consistency, measurements are usually taken from vessel branches with the most pronounced WSS. In patients with aneurysms, this typically meant the branch where the aneurysm developed or—when the aneurysm spanned multiple branches or the apex—the branch showing the highest WSS. In control subjects, the same high-WSS branch selection rule was applied. The directionality variable (dirWSSG) is categorized based on whether the WSSG vector followed the same path as the WSS vector from the same point. When the two vectors aligned, the case was labeled dirWSSG$^+$; when they opposed, it was marked dirWSSG$^-$.

$$WSSG = \sqrt{\left(\frac{\partial \tau_{w,p}}{\partial p}\right)^2 + \left(\frac{\partial \tau_{w,q}}{\partial q}\right)^2} \qquad \text{(Eq. 2)}$$

In this context, $\tau_w$ represents the WSS vector. The p-direction denotes the dominant or time-averaged orientation of WSS across the cardiac cycle, while the q-direction is defined as orthogonal to this primary flow direction.

Elevated WSSG is linked to faster blood flow and greater mechanical load on the aneurysm wall. However, there is ongoing debate regarding whether high or low WSS contributes more to rupture risk. Meng et al. [64] proposed that both extremes—high and low WSS—can promote aneurysm progression and rupture through distinct mechanisms. Specifically, low WSS combined with high OSI tends to initiate inflammation-driven degenerative remodeling, while high WSS along with positive WSSG promotes wall degeneration via mural cell activity. Dolan et al. [100] further noted that a negative WSSG may counteract the effects of high WSS on gene expression. Meng et al. [64] proposed a unifying hypothesis where both high and low WSS contribute to rupture: low WSS with high oscillatory shear index (OSI) promotes inflammatory remodeling, while high WSS with positive WSSG triggers mural cell-driven remodeling. Dolan et al. [100, 112] also found that positive WSSG increases genes promoting proliferation and ECM degradation (e.g., ADAMTS1), while downregulating anti-inflammatory genes and inhibitors of apoptosis. Negative WSSG may counteract these effects even under high WSS. Zimny et al. [111] conducted a case-control CFD study involving 38 patients with unruptured MCA aneurysms and 39 non-aneurysmal controls, using velocity profiles from transcranial color-coded sonography. They found that positive WSSG was the only significant independent predictor of aneurysm formation, particularly near bifurcation apices with coexisting high WSS.

### 3.3. The Pressure Difference (PD)

Endovascular coil embolization is widely regarded as a safe and effective approach for treating intracranial aneurysms [113, 114]. Nonetheless, compared to surgical clipping, this method shows a higher likelihood of aneurysm recurrence [115-118]. Understanding and evaluating the recurrence risk following coil embolization has become a critical area of focus [119, 120]. Nambu et al. [121] explored the hemodynamic predictors of post-coiling aneurysm recurrence through CFD simulations utilizing a virtual post-coiling model (VM). This model was generated by processing pre-treatment data and artificially segmenting the aneurysm dome with a flat plane that separates it from the parent artery or branching vessel. Their findings indicated that among various morphological and hemodynamic factors, PD was the most robust predictor of recurrence.

PD is a hemodynamic parameter used to identify regions potentially associated with thin aneurysm walls. It is calculated by first determining the difference between the local pressure and the average pressure across the entire computational domain. This pressure difference is then normalized by the dynamic pressure on the inlet. The resulting nondimensional PD value allows for consistent comparison across different cases by eliminating dependency on absolute pressure values and accounting for flow intensity.

$$PD = \frac{Pressure - Pressure_{ave}}{\frac{1}{2}\rho V_{in}^2} \quad \text{(Eq. 3)}$$

Suzuki et al. [122] showed that the PD is a useful hemodynamic indicator for identifying thin-walled regions (TWRs) in unruptured cerebral aneurysms. High PD values were found to correspond with areas likely to be structurally weaker, suggesting their potential for predicting rupture-prone sites. Furthermore, Uno et al. [123] analyzed 50 internal carotid artery aneurysms using CFD and found that a high pressure difference (PD) at the coil mass surface was strongly associated with aneurysm recurrence. Using both a virtual model (VM) and a real post-coiling model (RM), they showed that PD in RM had excellent predictive power (AUC = 0.977), confirming PD as a reliable, noninvasive indicator of recurrence risk after coil embolization.

### 3.4. Wall Shear Stress Divergence WSSD

WSS is a key hemodynamic indicator acting on the aneurysm wall. To account for both its directional components and spatial variations, WSSD was introduced, as defined:

$$WSSD = \frac{\partial WSS_x}{\partial x} + \frac{\partial WSS_y}{\partial y} + \frac{\partial WSS_z}{\partial z} \quad \text{(Eq. 4)}$$

Tanaka et al. [124] demonstrated that regions with high WSSD often overlapped with thin-walled areas of cerebral aneurysms, indicating that the maximum WSSD ($WSSD_{max}$) can serve as a useful marker for identifying structurally vulnerable regions. To enable comparison of WSSD values across different patients, the parameter is nondimensionalized. This normalization ensures that variations due to individual differences in geometry or flow conditions are minimized, allowing for a more consistent assessment of hemodynamic effects. To nondimensionalize WSSD, the components of wall shear stress ($WSS_x$, $WSS_y$, $WSS_z$) were each divided by the dynamic pressure measured at the inlet plane. This process yields the normalized components $WSS_x$, $WSS_y$ and $WSS_z$, which are then used to compute the nondimensionalized WSSD.

$$WSS_x^* = \frac{WSS_x}{\frac{1}{2}\rho v_{in}^2}, \quad WSS_y^* = \frac{WSS_y}{\frac{1}{2}\rho v_{in}^2}, \quad WSS_z^* = \frac{WSS_z}{\frac{1}{2}\rho v_{in}^2} \quad \text{(Eq. 5)}$$

Higher nondimensionalized WSSD (WSSD*) values indicate regions where the aneurysm wall experiences significant stretching due to divergent blood flow—such as when the flow strikes the wall and disperses radially or travels along the surface, creating outward tension. Conversely, lower WSSD* values suggest compressive forces acting on the aneurysm wall, which may reflect areas under inward-directed stress.

Kim et al. [125] found that WSSD showed the highest correspondence (86.67%) with thin-walled regions in unruptured MCA aneurysms, suggesting it may serve as a reliable indicator of structurally weak, rupture-prone areas.

### 3.5. Temporal Wall Shear Stress Gradient (TWSSG)

The temporal wall shear stress gradient (TWSSG) quantifies how rapidly the magnitude of wall shear stress (WSS) changes over time during a cardiac cycle. It is defined as:

$$TWSSG = \frac{\partial \tau_w}{\partial t} \quad \text{(Eq. 6)}$$

TWSSG may contribute to aneurysm initiation. While its magnitude remained unchanged with rotated branches, higher TWSSG was reported to be linked to morphological features associated with aneurysms, particularly in geometries with area variation [126]. White et al. [127] reported that the endothelium is sensitive to TWSSG, indicating that temporal fluctuations in shear stress may influence endothelial behavior and contribute to aneurysm initiation.
Atherosclerotic plaques have been reported to be closely associated with low TAWSS [128-130] and high TWSSG [131-133], highlighting the role of both sustained low shear and temporal fluctuations in plaque development.

### 3.6. Gradient of TAWSS (TAWSSG)

The gradient of time-averaged wall shear stress (TAWSSG) is defined according to the method proposed by Meng and colleagues, based on the framework introduced by [112, 134]. This approach distinguishes between positive and negative gradients in relation to the primary shear direction, denoted as $\hat{p}$. Unit vectors were introduced to represent the orientation of the TAWSS vector and its perpendicular direction. These are denoted as $\hat{p}$ for the primary (aligned) direction and $\hat{q}$ for the orthogonal (perpendicular) direction.

$$\hat{p} = \frac{\int_0^T \tau_w dt}{\left|\int_0^T \tau_w dt\right|}, \quad \hat{q} = \hat{p} \times \hat{n}$$
$$TAWSSG = \nabla_S(TAWSS).\hat{p} \quad \text{(Eq. 7)}$$

Here, $\nabla_S$ denotes the surface gradient operator, which is applied along the tangential plane of the vessel wall. The concept of disturbed flow arises when the direction of WSS varies significantly throughout the cardiac cycle, no longer maintaining consistent alignment with the primary vector $\hat{p}$. This fluctuating behavior reflects complex, multidirectional shear environments. To capture this phenomenon, transverse WSS (transWSS) was introduced as a metric that isolates the components of shear acting perpendicular to the dominant flow direction. Originally developed by [135] in the context of atherosclerosis research, transWSS has since become a useful tool for quantifying flow irregularity in vascular studies.

### 3.7. Transverse WSS (transWSS)

Transverse wall shear stress (transWSS) quantifies the component of WSS that acts perpendicular to the dominant shear direction. It is calculated as the time-averaged magnitude of the WSS vector projected along the $\hat{q}$ direction—orthogonal to the primary flow axis.

$$transWSS = \frac{1}{T}\int_0^T |\tau_w.\hat{q}| \, dt \quad \text{(Eq. 8)}$$

### 3.8. WSS pulsatility index (WSSPI)

To assess how the magnitude of WSS fluctuates over the course of a cardiac cycle, the WSSPI, originally introduced by Gosling et al. [136] is used as a quantitative measure of temporal variation.

$$WSSPI = \frac{\max \tau_w - \min \tau_w}{TAWSS}, \quad t \in [0, T] \quad \text{(Eq. 9)}$$

Interpretation of WSS magnitudes requires caution, as values can differ considerably depending on whether patient-specific or estimated boundary conditions are used in CFD simulations [137-140]. To address this variability, the distribution of TAWSS sometimes is analyzed using a normalized approach [141], by dividing TAWSS by the space-averaged TAWSS across the vessel branch (TAWSSB). Previous studies have demonstrated that, particularly in aneurysm models, this normalization yields WSS distributions that remain stable despite variations in boundary condition assumptions [138]. Similarly, both TAWSSG and transWSS were normalized by TAWSSB.

### 3.9. Oscillatory shear index (OSI)

The Oscillatory Shear Index (OSI) is another key hemodynamic indicator, used to assess the degree of directional variation in wall shear stress over the cardiac cycle [24, 49, 64, 142, 143]. Unlike TAWSS, which captures magnitude, OSI focuses on the extent to which shear stress reverses direction, a hallmark of disturbed and unstable blood flow. Elevated OSI levels have been linked to regions prone to aneurysm formation and expansion. In such environments, endothelial cells tend to shift toward a synthetic and pro-inflammatory state, initiating matrix degradation and weakening the vessel wall, biological changes that increase the likelihood of aneurysm progression.

$$OSI = \frac{1}{2}\left\{1 - \frac{\left|\int_0^T WSS_i dt\right|}{\int_0^T |WSS_i| dt}\right\} \quad \text{(Eq. 10)}$$

### 3.10. Aneurysm formation indicator (AFI)

Endothelial cells are believed to respond most strongly to temporal variations in WSS, prompting the development of several models aimed at quantifying these fluctuations [144-146]. A common feature across these models is the use of a reference vector, typically the time-averaged WSS vector, based on the idea that endothelial alignment tends to follow the dominant direction of shear over time. Among the various metrics proposed, the OSI, originally introduced by [146], is the most widely applied. OSI is calculated using the instantaneous WSS vector and captures the degree of directional change in shear stress throughout the cardiac cycle. While OSI provides a general measure of directional variability in WSS, it does not capture the instantaneous relationship between WSS fluctuations and endothelial response at specific time points in the cardiac cycle. To address this limitation, the Aneurysm Formation Indicator (AFI) was proposed as a more localized and temporally sensitive metric [147]. Building on the assumption that endothelial cells align with the dominant shear direction, AFI quantifies how much the instantaneous WSS vector deviates from the time-averaged direction by calculating the cosine of the angle between them. This approach was motivated by observations in pre-aneurysmal arterial segments, where both WSS magnitude remained low, and its direction changed significantly over the cycle. By focusing on these directional shifts at critical moments, such as midsystolic deceleration, AFI aims to pinpoint regions with disturbed flow patterns that may predispose vessels to aneurysm formation. Although still preliminary, AFI offers a refined perspective on WSS dynamics that complement existing models like OSI. AFI quantifies this deviation by computing the cosine of the angle between the instantaneous and time-averaged WSS vectors.

$$\cos(\theta) = \frac{WSS_i * WSS_{avg}}{|WSS_i| * |WSS_{avg}|} \quad \text{(Eq. 11)}$$

## 3.11. Oscillation velocity index (OVI)

Blood velocity reflects the local magnitude of flow at a given point, while the Oscillatory Velocity Index (OVI) captures variations in both direction and intensity of velocity throughout the cardiac cycle [148]. In addition, vorticity characterizes the rotational behavior of flow, offering insight into vortex formation based on changes in velocity fields.

$$OVI = \frac{1}{2}\left(1 - \frac{\left|\int_0^T fv_i \, dt\right|}{\int_0^T |fv_i| \, dt}\right) \quad \text{(Eq. 12)}$$

In this context, $fv_i$ denotes the instantaneous flow velocity vector, while $T$ represents the duration of a single cardiac cycle.

Tanioka et al. [149] analyzed 129 cerebral aneurysms and found that a high OVI was significantly associated with ruptured aneurysms, irregular morphology, and unstable flow patterns. High OVI correlated with higher WSS, WSS gradient, OSI, and complex geometry, suggesting that OVI may serve as a useful marker for assessing rupture risk.

## 3.12. Gradient oscillatory number (GON)

GON introduced by Shimogonya et al. [150], serves as a potential hemodynamic marker for identifying regions prone to cerebral aneurysm formation. This dimensionless index, ranging from 0 to 1, captures the extent of temporal variability in the spatial gradients of wall shear stress across the cardiac cycle. By measuring the oscillation of tangential force directions on the vessel wall during pulsatile flow, GON offers insight into disturbed shear environments associated with aneurysm initiation.

$$GON = 1 - \frac{\left|\int_0^T G \, dt\right|}{\int_0^T |G| \, dt} \quad (0 \leq GON \leq 1) \quad \text{(Eq. 13)}$$

where T is the period of the cardiac cycle and G is the spatial wall shear stress gradient vector.

## 3.13. Relative Residence Time (RRT)

RRT quantifies how long particles remain near the vessel wall, offering insight into regions of prolonged flow stagnation [151]:

$$RRT = \frac{1}{(1 - 2 \cdot OSI) \cdot TAWSS} \quad \text{(Eq. 14)}$$

RRT plays a key role in aneurysm hemodynamics, as prolonged particle residence near the wall is associated with an increased likelihood of thrombus formation and potential aneurysm occlusion [152]. Elevated RRT values indicate regions of blood stasis, which are linked to a higher likelihood of thrombus development [58].

Table 3: Summary of WSS, OSI, and RRT Assessment in Cerebral Aneurysm Studies

| Author & year | Hemodynamics parameter | Investigation of Hemodynamic Behavior in Cerebral Aneurysms |
|---|---|---|
| Bozorgpour et al., (2025) [153] | WSS, OSI, RRT, ECAP | High WSS, low OSI, low RRT and low ECAP lead to aneurysm rapture. |
| Zhu et al., (2023) [154]) | SR and OSI | may have predictive values for the risk of intracranial aneurysm rupture |
| Zimny et al., (2021) [111] | WSS and WSSG | IA development is driven by hemodynamic factors; high WSS influences MCA aneurysm formation, with positive WSSG playing a key promoting role. |
| Jiang et al., (2019) [155] | WSS | Low WSS can lead to thin aneurysm wall regions and may promote wall remodeling through blood–endothelial cell interaction. |
| Staarmann et al., (2019) [22] | WSS | Brief review of WSS in cerebral aneurysm formation, progression, rupture, and wall remodeling. |
| Tanioka et al., (2019) [149] | OVI | High OVI is linked to ruptured aneurysms, unstable flow, and complex geometry, making it a potential rupture risk marker. |
| Wang et al., (2018) [156] | High and Low WSS | Aneurysms grow under low WSS but rupture under high WSS conditions. |
| Zhang et al., (2016) [23] | Low WSS and high OSI | Low WSS and high OSI mark rupture sites in aneurysms, typically without impingement flow. |
| Sugiyama et al., (2016) [157] | RRT | Residence time near the aneurysm wall was estimated using the age-of-fluid method; maximum RRT appeared only in atherosclerotic aneurysms. |
| H. Meng et al., (2014) [64] | Low WSS with OSI | Cell-mediated inflammation is linked to the progressive growth and rupture of aneurysms. |
| H. Meng et al., (2014) [64] | High WSS | Triggered wall cell signaling, promoting formation and rupture of primary or secondary bleb aneurysms. |
| Dolan et al., (2013) [100] | High WSS and Positive WSSG | High WSS and positive WSSG are linked to aneurysm initiation, plaque stabilization, and endothelial damage. |
| Dolan et al., (2013) [100] | High and Low WSS | WSS above 3 Pa is considered high, while below 1 Pa is considered low. |
| Sugiyama et al., (2013) [158] | RRT | High RRT is centered in vortex regions, where prolonged blood circulation is linked to intra-aneurysmal atherosclerotic lesions. |
| Xiang et al. (2011) [58] | WSS, OSI, and RRT | Threshold values of hemodynamic parameters were used to distinguish ruptured from unruptured aneurysms. |

### 3.14. Endothelial Cell Activation Potential (ECAP)

ECAP introduced by Di Achille et al. [159], identifies regions exposed to a combination of low TAWSS and elevated OSI. It is defined as:

$$ECAP = \frac{OSI}{TAWSS} \quad \text{(Eq. 15)}$$

ECAP is often used to assess the thrombogenic potential of the arterial wall, with elevated values indicating endothelial vulnerability. Regions exhibiting high OSI combined with low TAWSS tend to produce high ECAP values, which are linked to pro-thrombotic and dysfunction-prone conditions [159, 160].

### 3.15. Helicity

The helicity, $H(t)$, serves as a scalar metric to detect streamwise vortical patterns by measuring how well the velocity vector $v(x,t)$ aligns with the vorticity vector $\omega(x,t)$ within a given volume $V$. It is defined by the following volume integral:

$$H(t) = \int_V v(x,t) \cdot \omega(x,t) dv = \int_V H_k(x,t) dv \quad \text{(Eq. 16)}$$

Here, $H_k$ denotes the helicity density. The scalar helicity $H(t)$ can also be computed over a 2D surface by integrating the corresponding quantities across that plane. The sign of $H_k$ reveals the rotational direction with respect to the flow: a positive value suggests a right-handed (clockwise) helical structure, while a negative value corresponds to a left-handed (counterclockwise) configuration. Since net helicity of zero may arise from either opposing vortex pairs or a lack of flow rotation entirely, the absolute value of helicity serves as a useful metric to differentiate these cases:

$$|H(t)| = \int_V |H_k(x,t)| dv \quad \text{(Eq. 17)}$$

Helical flow is commonly observed in healthy aortic circulation and has been shown to mitigate flow irregularities in cerebral aneurysms. Since the helical patterns within the aorta typically extend beyond the thin boundary layer of the vessel wall, they are more effectively captured using 4D flow MRI (4DMR) compared to wall shear stress (WSS) assessments. Consequently, exploring the relationship between helicity and WSS is of interest due to the diagnostic potential and clinical relevance of WSS measurements. Local Normalized Helicity (LNH) is expressed as:

$$LNH(x,t) = \frac{H_k(x,t)}{|v(x,t)||\omega(x,t)|} \quad \text{(Eq. 18)}$$

LNH is frequently employed to depict vortical formations in the aorta by rendering isosurfaces at symmetric but opposite values—commonly at ±0.6. To quantitatively evaluate helicity, the helicity density $H_k$ can be averaged across a specified spatial volume $V$ and temporal duration $T$, using the following expressions:

$$\begin{aligned} h_1 &= \frac{1}{TV} \int_T \int_V H_k dV dt \\ h_2 &= \frac{1}{TV} \int_T \int_V |H_k| dV dt \\ h_3 &= \frac{h_1}{h_2} \end{aligned} \quad \text{(Eq. 19)}$$

The quantity $h_1$ becomes zero when the helical structures exhibit mirror symmetry or when both velocity and vorticity are absent. In contrast, $h_2$ captures the overall helicity magnitude within the volume, independent of rotational direction. The ratio $h_3$, which ranges from –1 to 1, indicates the dominance and balance between right- and left-handed helicity, with its sign revealing the prevailing rotational sense.

### 3.16. Flow Complexity parameters

Hodis [161, 162] developed an analytic expression to quantify flow complexity in cerebral aneurysms and examined its link to rupture status using five ruptured and five unruptured anterior communicating artery cases. They computed the parameter in jet and non-jet regions, finding that ruptured aneurysms had jets that were ~4.5 times more concentrated (less complex) and non-jet regions that were ~3.5 times more complex than in unruptured cases. A strong positive correlation was also found between non-jet complexity and dome volume in ruptured aneurysms. This kinematic parameter provides a quantitative measure of flow complexity that may improve rupture risk assessment.

The proposed a kinematic-based method to quantify flow complexity using local curvature $k$ and torsion $\tau$ of the velocity field, defined as:

$$k = \sqrt{(\frac{\omega}{v})^2 - (\frac{\overline{\omega v}}{v^2})^2}, \quad \tau = \frac{\overline{\omega v}}{v^2} \quad \text{(Eq. 20)}$$

where $\vec{\omega}$ and $\vec{v}$ are the velocity and vorticity vectors. These quantities capture rotational features and local complexity without the need to extract flow lines. Large values of curvature and torsion correspond to complex flow regions. To derive a single flow complexity metric, the authors combined these parameters using the Fundamental Theorem of Curves [163].

### 3.17. Low WSS Area Ratio (LSAR)

LSAR quantifies the portion of the aneurysm wall where wall shear stress falls below 10% of the mean WSS in the parent artery, normalized by the aneurysm dome area [68]. Jiang et al. [164] demonstrated that thick-walled atherosclerotic aneurysms are linked to low WSS and higher LSAR.

$$LSAR = \frac{Low\ wall\ shear\ stress\ area}{Aneurysm\ dome\ area} \quad \text{(Eq. 21)}$$

### 3.18. Aneurysm Number (An)

An represents the ratio between transport and vortex formation time scales in IAs, providing insight into flow behavior within the aneurysm sac [165]. It distinguishes whether vortex structures can develop and persist within the dome before being carried out by the parent flow. In sidewall aneurysms, the transport time is defined as the time required for flow to move a fluid particle across the neck, while in bifurcation aneurysms, it reflects the time needed to transport a particle from the inlet to the outlets. Based on this, An is formulated using a geometric parameter (neck or outlet diameter), a scaling constant $\alpha$ (1 for sidewall and 2 for bifurcation IAs), and the pulsatility index (PI).

$$An = \alpha \frac{W_{mod}}{D} PI, \quad PI = \frac{PSV - EDV}{U} \quad \text{(Eq. 22)}$$

When $An > 1$, the vortex formation occurs faster than fluid transport, allowing a vortex to establish within the aneurysm dome. Conversely, $An < 1$ indicates that the flow exits the aneurysm before a vortex can form, resulting in a more stable shear layer across the sac. This parameter helps characterize intra-aneurysmal flow regimes, which may relate to aneurysm stability and rupture risk. Here, PSV refers to the peak systolic velocity, EDV is the end diastolic velocity, and U denotes the time-averaged velocity over a complete cardiac cycle.

### 3.19. Q-criteria

Vortex structures play a significant role in aneurysmal flow and can be described using the Q-criterion [166, 167]. This criterion is derived from the velocity gradient tensor, which is expressed as the sum of the rate of strain tensor (S) and the vorticity tensor ($\Omega$). The strain tensor S is defined as the symmetric part of the velocity gradient, while $\Omega$ represents its antisymmetric part.

$$\nabla \vec{u} = S + \Omega$$
$$S = \frac{1}{2}[(\nabla \vec{u}) + (\nabla \vec{u})^T], \ \Omega = \frac{1}{2}[(\nabla \vec{u}) - (\nabla \vec{u})^T] \quad \text{(Eq. 23)}$$

According to Hunt et al.[168], a vortex is identified in regions where:

$$Q = \frac{1}{2}[|\Omega|^2 - |S|^2] > 0 \quad \text{(Eq. 24)}$$

The Q-criterion identifies regions where the vorticity tensor outweighs the strain tensor.

### 3.20. Inflow Concentration Index (ICI)

ICI [166] quantifies how focused the incoming blood flow is at the aneurysm entrance. It is calculated as the ratio of the percentage of parent artery flow entering the aneurysm to the percentage of the ostium area exhibiting positive inflow velocity.

$$ICI = \frac{Q_{in}/Q_v}{A_{in}/A_o} \quad \text{(Eq. 25)}$$

Here, $Q_{in}$ represents the inflow rate into the aneurysm, $Q_v$ is the flow rate in the parent artery, $A_{in}$ denotes the area of the region with positive inflow velocity, and $A_o$ is the total area of the aneurysm ostium.

### 3.21. Shear Concentration Index (SCI)

SCI [166] quantifies how localized the WSS is within the aneurysm sac. Specifically, it compares the distribution of shear forces in regions with elevated WSS—defined as areas where WSS exceeds the mean WSS of the adjacent vessel wall by more than one standard deviation—to the entire sac. Let $A_h$ denote the area with high WSS and $A_a$ the total sac area. The corresponding shear forces over these regions are $F_h$ and $F_a$, respectively, calculated as:

$$F_h = \int_{A_h} |\tau| dA, \quad F_h = \int_{A_a} |\tau| dA \quad \text{(Eq. 26)}$$

The SCI is then defined as the ratio of normalized shear force to normalized area:

$$SCI = \frac{F_h/F_a}{A_h/A_a} \quad \text{(Eq. 27)}$$

A higher SCI suggests a more concentrated distribution of WSS within localized regions of the aneurysm wall.

### 3.22. Low Shear Area (LSA)

LSA [166] quantifies the portion of the aneurysm sac exposed to abnormally low WSS. It is calculated as the ratio of the area within the aneurysm where WSS falls below one standard deviation below the mean WSS of the parent artery to the total sac area. Denoting $A_l$ as the area under low WSS and $A_a$ as the total aneurysm sac area, LSA is expressed as:

$$LSA = A_l/A_a \quad \text{(Eq. 28)}$$

An LSA value close to 1 indicates that most of the sac is subjected to low WSS, whereas a value near 0 suggests minimal exposure to such low shear conditions.

### 3.23. Low Shear Index (LSI)

LSI [166] captures the proportion of total shear force distributed over regions experiencing abnormally low WSS. It is calculated by multiplying the normalized shear force within the low-WSS region by the corresponding area ratio. Specifically:

$$LSI = \frac{F_l \cdot A_l}{F_a A_a} \quad \text{(Eq. 29)}$$

where $F_l$ is the total shear force in the low-WSS region $A_l$, and $F_a$, $A_a$ are the total shear force and area of the entire aneurysm sac, respectively:

$$F_l = \int_{A_l} |\tau| dA \quad \text{(Eq. 30)}$$

The LSI ranges from 0 to 1, where 0 indicates no shear force is applied in low-WSS areas, and 1 indicates all shear force is concentrated in those regions.

### 3.24. Kinetic Energy Ratio (KER) and Viscous Dissipation Ratio (VDR)

KER [166] assesses the relative kinetic energy contained within the aneurysm compared to that in the adjacent parent artery segment. It is computed as the volume-averaged kinetic energy within the aneurysm sac divided by the volume-averaged kinetic energy in the near-vessel region:

$$KER = \frac{\int_{V_a} \frac{1}{2} u^2 \, dV / V_a}{\int_{V_{near}} \frac{1}{2} u^2 \, dV / V_{near}} \quad \text{(Eq. 31)}$$

where $u$ is the velocity magnitude, $V_a$ is the aneurysm volume, and $V_{near}$ is the volume of the adjacent parent vessel.

VDR quantifies the relative rate of mechanical energy dissipation due to viscosity in the aneurysm sac compared to the near-parent artery. It is defined as:

$$VDS = \frac{\int_{V_a} 2\mu/\rho (e_{ij} e_{ij}) \, dV / V_a}{\int_{V_{near}} 2\mu/\rho (e_{ij} e_{ij}) \, dV / V_{near}} \quad \text{(Eq. 32)}$$

Here, $\mu$ is the dynamic viscosity, $\rho$ is the fluid density, and $e_{ij}$ is the strain rate tensor, calculated as:

$$e_{ij} = \frac{1}{2} \left( \frac{\partial u_i}{\partial x_j} + \frac{\partial u_j}{\partial x_i} \right) \quad \text{(Eq. 33)}$$

Higher values of VDR suggest greater viscous energy loss within the aneurysm compared to the parent artery.

### 3.25. Turbulent Kinetic Energy (TKE)

Turbulent Kinetic Energy (TKE) represents the kinetic energy per unit mass associated with velocity fluctuations $u'_i$ in turbulent flow and is expressed in units of m²/s² [169].

$$k = \frac{1}{2} \overline{u'_i u'_i} \quad \text{(Eq. 34)}$$

Where $u'_i$ is a turbulent fluctuation. TKE reflects the intensity of velocity fluctuations in blood flow and can serve as an indicator of disturbed hemodynamics within cerebral aneurysms. In this context, elevated TKE may signify regions of complex, unstable flow that are associated with adverse mechanical forces on the aneurysm wall. Such conditions can contribute to pathological wall remodeling or degeneration. Therefore, assessing TKE provides a quantitative means to evaluate intraluminal flow disturbance, potentially helping to identify aneurysms at higher risk of rupture.

### 3.26. Fluctuating Kinetic Energy (FKE)

To evaluate the FKE [170], the instantaneous velocity $u_i(x,t)$ is decomposed into a mean component $\bar{u}_i(x,t)$ and a fluctuating component $u'_i(x,t)$. The FKE is then defined as the time-averaged kinetic energy associated with these velocity fluctuations:

$$FKE(x,t) = \frac{1}{2}\langle u'_i(x,t) \cdot u'_i(x,t) \rangle \quad \text{(Eq. 35)}$$

FKE quantifies cycle-to-cycle variations in turbulent flow and serves as an indicator of the level of turbulent activity. Huang et al. [171] found FKE was substantially higher in ruptured aneurysms, particularly when using non-Newtonian models, with up to a 30% increase compared to Newtonian simulations. In contrast, unruptured aneurysms showed minimal flow fluctuations and little difference between fluid models, suggesting FKE may be a key indicator of rupture risk. Xu et al. [170] conducted CFD simulations on two matched pairs of ruptured and unruptured aneurysms and found that ruptured aneurysms exhibited highly disturbed flow, with notable velocity and WSS fluctuations, especially during late systole. These aneurysms also showed obvious intra-cycle and cycle-to-cycle WSS fluctuations, suggesting that flow instability may be linked to aneurysm rupture. Varble et al. [172] analyzed 56 MCA aneurysms using high-resolution CFD and found that FKE did not distinguish ruptured from unruptured aneurysms. However, they observed a positive correlation between FKE and aneurysm size and size ratio, suggesting that flow instability may be related to inflow jet breakdown, particularly in larger aneurysms.

## 4. Conclusion

Understanding the hemodynamic environment within cerebral aneurysms is essential for elucidating the mechanisms of aneurysm initiation, progression, and rupture. Over the past two decades, a wide array of hemodynamic variables—ranging from basic metrics like wall shear stress (WSS) and oscillatory shear index (OSI) to more advanced descriptors such as relative residence time (RRT), wall shear stress gradient (WSSG), and endothelial cell activation potential (ECAP)—have been investigated using computational and imaging-based techniques. These variables capture different aspects of the flow environment, including the magnitude, directionality, spatial gradients, and temporal characteristics of the forces acting on the vessel wall.

However, the diversity of variables across studies has created a fragmented landscape, with each study often focusing on a limited subset of metrics tailored to its own hypotheses or modeling capabilities. This inconsistency not only limits the ability to compare findings across studies but also presents challenges for new researchers trying to determine which variables are most relevant for their objectives. Furthermore, the physiological significance and interpretive value of some variables—especially newer or compound metrics—are still being established, and their relationship to biological endpoints such as wall remodeling, inflammation, or rupture risk remains an active area of investigation.

To address these issues, this paper provides a consolidated reference of all hemodynamic variables commonly examined in cerebral aneurysm research. By systematically collecting and categorizing these variables based on their biomechanical basis and usage in literature, we offer a structured overview that can inform future study design. This catalog is not meant to endorse a particular set of metrics as universally superior, but rather to provide a comprehensive map of what has been done, what is frequently used, and where there may be gaps or opportunities for further exploration.

Ultimately, we hope this work serves as a foundational resource for both researchers and clinicians seeking to apply hemodynamic analysis to the study of cerebral aneurysms. Whether designing new computational models, interpreting imaging-derived metrics, or comparing results across studies, having a centralized reference of available variables can facilitate more consistent methodologies, improve reproducibility, and support the development of standardized approaches for hemodynamic-based risk assessment. As the field continues to evolve, especially with the integration of machine learning and patient-specific modeling, a shared understanding of hemodynamic metrics will be increasingly important for translating computational insights into clinical decision-making.

**Conflict of Interest**

The authors declare no conflict of interest.

**Ethical Approval**

This article does not contain any studies with human participants or animals performed by any of the authors.

**Funding**

This research received no specific grant from any funding agency in the public, commercial, or not-for-profit sectors.

**Data Availability**

This is a review article and does not involve the generation of new data. All documents and sources referenced in this study are publicly available and properly cited within the manuscript.

149. Tanioka, S., et al., *Quantification of hemodynamic irregularity using oscillatory velocity index in the associations with the rupture status of cerebral aneurysms.* Journal of NeuroInterventional Surgery, 2019. 11(6): p. 614-617.
150. Shimogonya, Y., et al., *Can temporal fluctuation in spatial wall shear stress gradient initiate a cerebral aneurysm? A proposed novel hemodynamic index, the gradient oscillatory number (GON).* Journal of biomechanics, 2009. 42(4): p. 550-554.
151. Himburg, H.A., et al., *Spatial comparison between wall shear stress measures and porcine arterial endothelial permeability.* American Journal of Physiology-Heart and Circulatory Physiology, 2004. 286(5): p. H1916-H1922.
152. Rayz, V., et al., *Flow residence time and regions of intraluminal thrombus deposition in intracranial aneurysms.* Annals of biomedical engineering, 2010. 38: p. 3058-3069.
153. Bozorgpour, R. and J.R. Rammer, *Hemodynamic Markers: CFD-Based Prediction of Cerebral Aneurysm Rupture Risk.* arXiv preprint arXiv:2504.10524, 2025.
154. Zhu, Y., et al., *Assessing the risk of intracranial aneurysm rupture using computational fluid dynamics: a pilot study.* Frontiers in Neurology, 2023. 14: p. 1277278.
155. Jiang, P., et al., *Hemodynamic characteristics associated with thinner regions of intracranial aneurysm wall.* Journal of Clinical Neuroscience, 2019. 67: p. 185-190.
156. Li, M., et al., *Hemodynamics in ruptured intracranial aneurysms with known rupture points.* World Neurosurgery, 2018. 118: p. e721-e726.
157. Sugiyama, S.-i., et al., *Computational Hemodynamic Analysis for the Diagnosis of Atherosclerotic Changes in Intracranial Aneurysms: A Proof-of-Concept Study Using 3 Cases Harboring Atherosclerotic and Nonatherosclerotic Aneurysms Simultaneously.* Computational and Mathematical Methods in Medicine, 2016. 2016(1): p. 2386031.
158. Sugiyama, S.-i., et al., *Relative residence time prolongation in intracranial aneurysms: a possible association with atherosclerosis.* Neurosurgery, 2013. 73(5): p. 767-776.
159. Di Achille, P., et al., *A haemodynamic predictor of intraluminal thrombus formation in abdominal aortic aneurysms.* Proceedings of the Royal Society A: Mathematical, Physical and Engineering Sciences, 2014. 470(2172): p. 20140163.
160. Boniforti, M.A., et al., *Image-based numerical investigation in an impending abdominal aneurysm rupture.* Fluids, 2022. 7(8): p. 269.
161. Hodis, S., *Correlation of flow complexity parameter with aneurysm rupture status.* International Journal for Numerical Methods in Biomedical Engineering, 2018. 34(11): p. e3131.
162. Hodis, S., D.F. Kallmes, and D. Dragomir-Daescu, *Adaptive grid generation in a patient-specific cerebral aneurysm.* Physical Review E—Statistical, Nonlinear, and Soft Matter Physics, 2013. 88(5): p. 052720.
163. Struik, D.J., *Lectures on classical differential geometry*. 2012: Courier Corporation.
164. Jiang, P., et al., *Hemodynamic findings associated with intraoperative appearances of intracranial aneurysms.* Neurosurgical review, 2020. 43: p. 203-209.
165. Asgharzadeh, H. and I. Borazjani, *A non-dimensional parameter for classification of the flow in intracranial aneurysms. I. Simplified geometries.* Physics of Fluids, 2019. 31(3).
166. Cebral, J.R., et al., *Quantitative characterization of the hemodynamic environment in ruptured and unruptured brain aneurysms.* American Journal of Neuroradiology, 2011. 32(1): p. 145-151.
167. Jiang, M., R. Machiraju, and D. Thompson, *14-detection and visualization of vortices.* Visualization handbook, 2005: p. 295-309.
168. Hunt, J.C., A.A. Wray, and P. Moin, *Eddies, streams, and convergence zones in turbulent flows.* Studying turbulence using numerical simulation databases, 2. Proceedings of the 1988 summer program, 1988.
169. Ha, H., et al., *Turbulent kinetic energy measurement using phase contrast MRI for estimating the post-stenotic pressure drop: in vitro validation and clinical application.* PloS one, 2016. 11(3): p. e0151540.
30